# Broadband imaging with one planar diffractive lens

Nabil Mohammad,[1] Monjurul Meem,[1] Bing Shen,[2] and Rajesh Menon[1,*]

[1] Department of Electrical and Computer Engineering, University of Utah, Salt Lake City UT 84112
[2] MACOM Technology Solutions, NY
* Correspondence to rmenon@eng.utah.edu

**We demonstrate imaging over the visible band using a single planar diffractive lens. This is enabled via multi-level diffractive optics that is designed to focus over a broad wavelength range, which we refer to as an achromatic diffractive lens (ADL). We designed, fabricated and characterized two ADLs with numerical apertures of 0.05 and 0.18. Diffraction-limited focusing is demonstrated for the NA=0.05 lens with measured focusing efficiency of over 40% across the entire visible spectrum (450nm to 750nm). We characterized the lenses with a monochromatic and a color CMOS sensor, and demonstrated video imaging under natural sunlight and other broadband illumination conditions. We use rigorous electromagnetic simulations to emphasize that ADLs can achieve high NA (0.9) and large operating bandwidth (300nm in the visible spectrum), a combination of metrics that have so far eluded other flat-lens technologies such as metalenses. These planar diffractive lenses can be cost-effectively manufactured over large areas and thereby, can enable the wide adoption of flat, low-cost lenses for a variety of imaging applications.**

Refractive lenses are bulky and often challenging to incorporate into imaging systems that have restricted form factors. Diffractive optics, on the other hand, may be planar and lightweight.[1] However, conventional diffractive optics are not typically used for imaging because of significant off-axis and chromatic aberrations[2,3] as well as low broadband focusing efficiencies. Recently, planar metalenses have been used for imaging.[4,5] Unfortunately, metalenses require subwavelength features and large aspect ratios, making them impractical for low-cost manufacturing over large areas. In addition, they usually suffer from polarization sensitivity[6-8] and possess significant chromatic aberrations.[9-13] Metalenses with conical wavefronts have been demonstrated for focusing.[14,15] However, these suffer from relatively low transmission efficiency, require very small features (~100nm) and therefore are difficult to scale to larger apertures. Here, we demonstrate that metalenses are not required for imaging light intensities, a scalar property of the electromagnetic field. Diffractive optics with super-wavelength features and relatively low aspect ratios, which are far simpler to fabricate, are sufficient for imaging light intensities. However, we note that metasurfaces are required to manipulate vector properties of light such as polarization.[16] We further emphasize that we previously demonstrated water-immersion diffractive lenses with numerical aperture (NA) as high as 1.43.[17]

Previously, we utilized the concept of broadband diffractive optics to design broadband holograms[18] and to demonstrate broadband spectrum splitting and concentration,[19,20] phase masks for 3D lithography[21] and cylindrical lenses with super-achromatic performance over the entire visible band.[22] Here, we design, fabricate and characterize broadband diffractive optics as planar lenses for imaging. Broadband operation is achieved by optimizing the phase transmission function for each wavelength carefully to achieve the desired intensity distribution at that wavelength in the focal plane. We designed these achromatic diffractive lenses (ADLs) by maximizing the focusing efficiency at the design wavelengths ranging from 450nm to 750nm in steps of 50nm. We numerically investigated multiple sampling schemes in wavelength and decided that our chosen scheme provided a good compromise between computational cost and average focusing efficiency. Each lens is comprised of concentric circular rings of width = 3μm and the height of each ring is varied between 0 and a maximum value (which was 2.4μm and 2.6μm for the NA=0.05 and 0.18 lenses, respectively). We designed and experimentally characterized two lenses each of focal length, f=1mm and numerical aperture (NA) of 0.05 and 0.18, respectively. All design parameters are summarized in the supplementary information (Table S1). Recently, broadband diffractive lenses have been applied to imaging.[23] However, due to very low resolution of these lenses, significant image blurring is observed, which requires significant post-processing to obtain sharp images. In contrast, our approach is able to maintain the quality of the images comparable to that achievable with more complex systems of lenses.

The diffractive lens can be accurately modeled by scalar diffraction theory in the regime of Fresnel approximation.[24,25] We utilized a modified version of direct-binary search to optimize the height profile of each lens. Details of the design process have been described elsewhere.[22] The on-axis focusing efficiency averaged over all the design wavelengths is used as the metric for optimization. Grayscale lithography was used to fabricate the lenses in a positive photoresist (Shipley 1813) spin coated on a glass wafer.[22] A photograph of one of the lenses is shown in Fig. 1(b) to confirm that this diffractive lens is indeed flat. Optical micrographs of the fabricated lenses are shown in Figs. 1(c) and 1(d), for NA=0.05 and 0.18, respectively. Figures 1(e) and 1(f) shows the measured full-width at half-maximum (FWHM) of the focal spot as a function of wavelength for NA=0.05 and 0.18 lenses, respectively. In Fig. 1(e), it is clear that the fabricated device is able to achieve diffraction-limited performance to within 5% of the diffraction-limited value. The NA=0.18 lens fails to achieve diffraction-limited spot size for some of the design wavelengths. Nevertheless, we show below that both lenses are capable of forming reasonably good quality images. In the case of the NA=0.18 lens, simple image processing such as deconvolution can significantly improve the imaging performance as well. The corresponding simulated focal spots (see supplementary figure S1) indicate that there is good agreement between experiments and simulations for both lenses. We note that the streaks that are visible in Fig. 1(c) arise from a stage calibration error in our patterning tool, which we believe has some effect on the experimental focusing efficiency as described later. These streaks can be eliminated with proper stage calibration. The diffraction rings seen in the focal spots of the NA=0.18 lens may be mitigated by the use of smaller ring widths and careful choice of the metric for optimization.

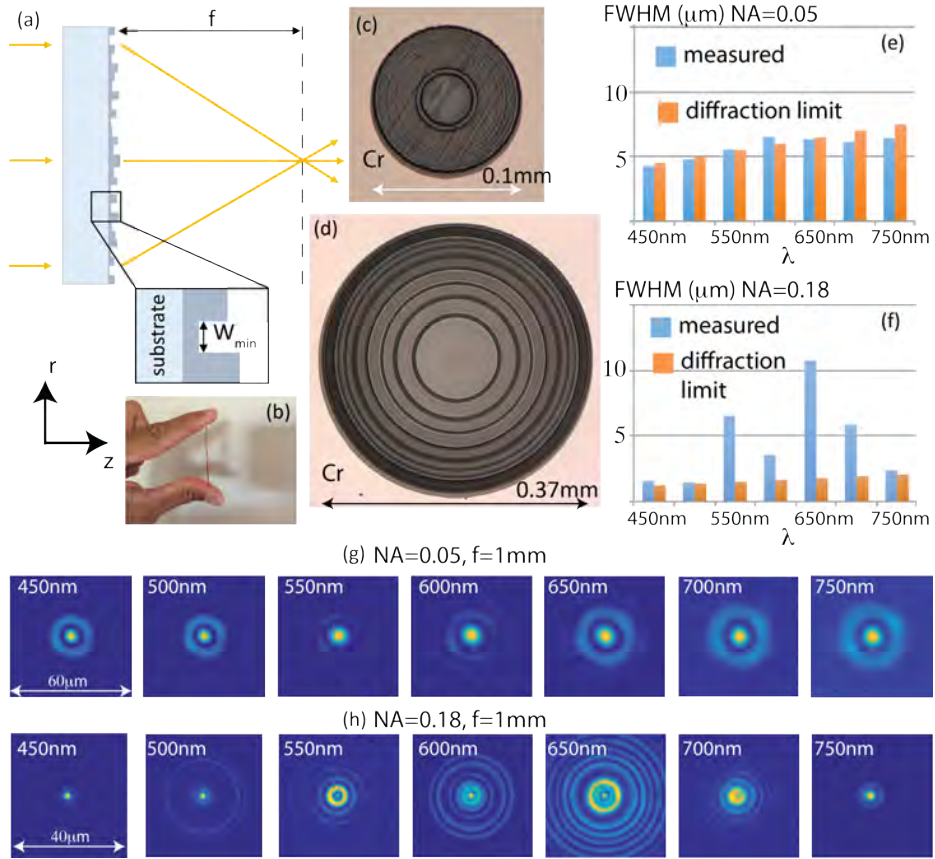

*Figure 1:* (a) Schematic of a flat-lens design. The structure is comprised of concentric rings of width, $W_{min}$ and varying heights. (b) Photograph of one fabricated lens. Optical micrographs of (c) NA=0.05 and (d) NA=0.18 lenses. Focal length is 1mm. Measured full-width at half-maximum (FWHM) of the focal spot as a function of wavelength for (e) NA=0.05 and (f) NA=0.18 lenses. Measured focal spots as a function of wavelength for (g) NA=0.05 and (h) NA=0.18 lenses.

A collimated beam from a super-continuum source equipped with a tunable filter illuminates the lens (see supplementary information).[22] The light distribution at the focus for each lens was then magnified by an objective-tube lens system and recorded on a monochrome sensor (see supplementary information for details) and are shown in Figs. 1(g) and (h) for NA=0.05 and 0.18 lenses, respectively.

The focusing efficiency, defined as the ratio of the integrated power over a circular aperture with diameter 18μm in the focal plane to the total power over the lens aperture as a function wavelength was also measured (Fig. 2a). Measurement details are included in the supplementary information. The simulated focusing efficiencies are included in the supplementary information (Fig. S2). Numerical analysis of the fabrication errors included in the supplementary information (see Fig. S3) seems to suggest that fabrication errors are one possible reason for the reduction in the measured focusing efficiencies. Nevertheless, the measured efficiency averaged over the 300nm bandwidth of 42% and 22.1% for NA=0.05 and 0.18 lenses, respectively, is considerably larger than the averaged efficiency of a comparable metalens over a much smaller bandwidth (60nm).[26] We also point out that the focusing efficiencies can be increased by allowing for smaller zone widths as discussed for the case of high-NA lenses later. We further note generally it is more challenging to design a lens with a larger aperture while maintaining the imaging performance. This might be mitigated in the future by combining refractive and diffractive power into a single aperture.

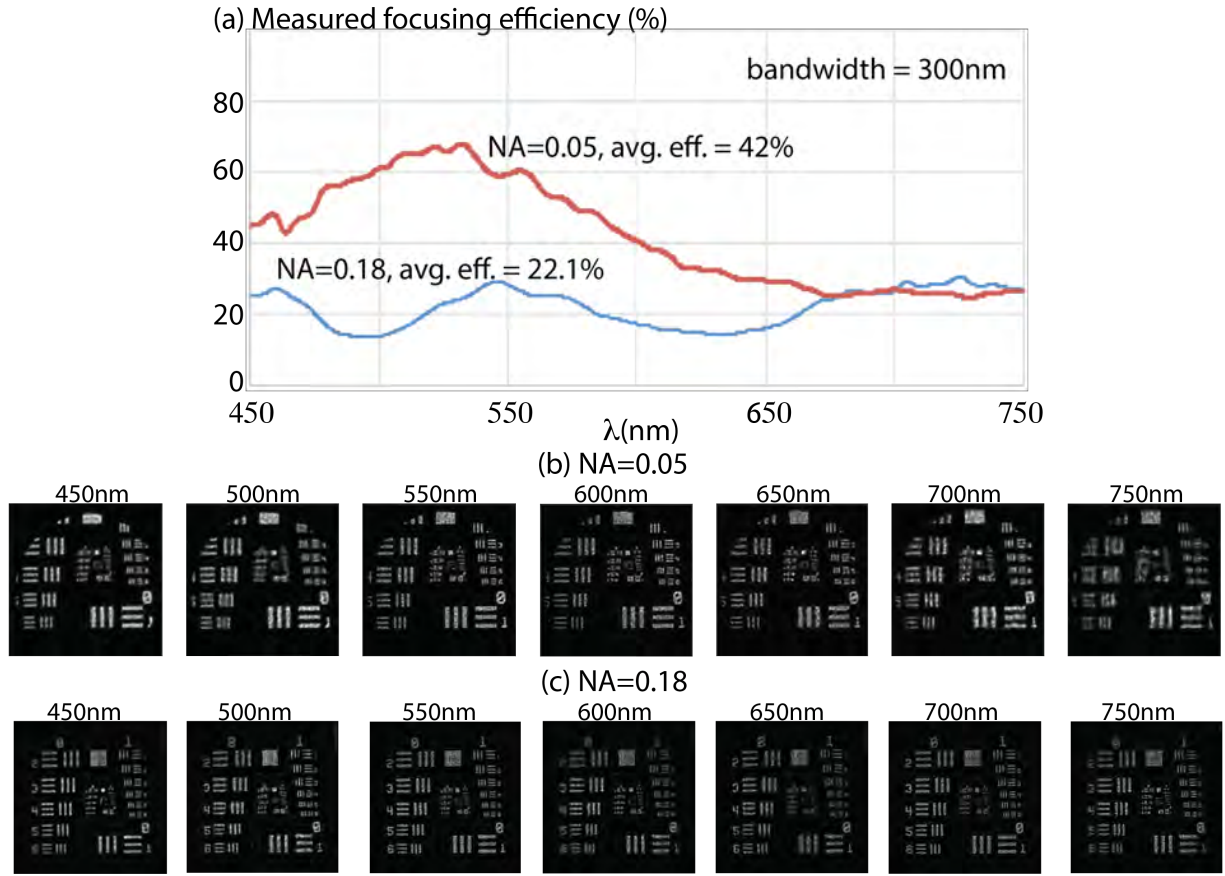

*Figure 2: (a) Measured focusing efficiency as a function of wavelength for the 2 lenses shown in Fig. 1. Images captured on a monochrome sensor of the Air Force resolution target in transmission at various illumination wavelengths for (b) NA=0.05 and (c) NA=0.18 lenses. Details of the experiments are in the text and in the supplementary information. Blind deconvolution was applied to these images.*

To demonstrate imaging capability, we used a standard test chart (USAF 1951) with a diffuser behind it as the object and imaged it onto the same monochrome sensor using each flat lens. Details of the experiment are described in the supplementary information. We used the same illumination source (varying center wavelength and bandwidth=10nm) as was used for focal spot characterization. A simple blind deconvolution method was applied to the raw images and the results are presented in Figs. 2(b) and 2(c) for NA=0.05 and 0.18 lenses, respectively. It is clear that our lenses are able to form relatively good images at all the wavelengths.

Finally, we built a color camera by placing the flat lens with a conventional CMOS sensor (DFM 72BUC02-ML, The Imaging Source). A variety of test images were captured using both artificial lighting as well under ambient sunlight. In the latter case, a conventional IR-cut filter was placed in front of the lens. The results are summarized in Fig. 3. Since our sensor can capture video data, we have also included two examples of videos as supplementary information for each of the two lenses.

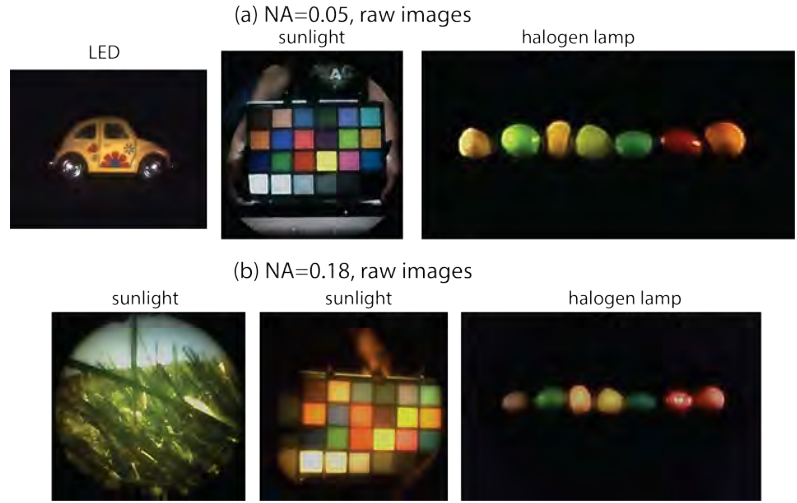

*Figure 3:* *Example photographs taken with a camera consisting of only a single flat lens and a conventional color CMOS sensor. An IR-cut filter is placed in front of the lens for images taken under sunlight. Video data taken with these cameras are included as supplementary material.*

As we have pointed out before, these diffractive lenses are also polarization-insensitive.[22] This ensures that the lenses perform equally well for any polarization input, which is important for general imaging applications. All experiments reported here were conducted with randomly polarized incident light.

In a diffractive lens, the maximum achievable NA is determined by $\lambda/(2*w)$, where w is the minimum feature size (constrained by fabrication). In order to achieve NA comparable to that of a mobile-phone camera (NA~0.3), one needs w ~0.83μm. We can expect the maximum pixel height to be similar to what we have used here. This means that for many photography applications, a diffractive lens would need an aspect ratio of only 1.7. This is considerably simpler to fabricate than a visible-wavelength metalens, where feature widths of <50nm and aspect ratios>15 are required.[5]

For ease of fabrication, we chose to demonstrate lenses with relatively low NA (large F/#). It is possible to overcome this limitation with improved fabrication processes. To illustrate this point, we designed a lens with NA=0.9 and f=3.5μm using minimum zone width of 0.25μm and maximum zone height of 1.25μm. Note that the maximum aspect ratio for this lens is 5:1 and the minimum feature is considerably larger than those required in metalenses. For ease of computation, we applied the same scalar diffraction model as before during the design process. The basic design methodology including the figure of merit was the same as for the low-NA lenses. Finally, we applied rigorous electromagnetic modeling via the finite-difference time-domain (FDTD) method to analyze the performance of the optimized design, which is shown in Fig. 4(a). The simulated focusing efficiency defined as the ratio of power inside 3 times full-width at half-maximum to that inside the lens aperture is plotted as a function of wavelength in Fig. 4(b), and the calculated average efficiency is 34% across an operating bandwidth of 300nm. The simulated focal spots at the design wavelengths are shown in Fig. 4(c). Although this preliminary design does not achieve diffraction-limited focusing at all the design wavelengths, these are sufficient for good image formation via post-processing using a-priori information of the focused spots as illustrated by the images formed by the fabricated NA=0.18 lens above.[23] Thereby, we confirm that our multilevel diffractive lens can achieve high-NA and large operating bandwidth, a combination of requirements that have so far eluded single-element imaging systems.

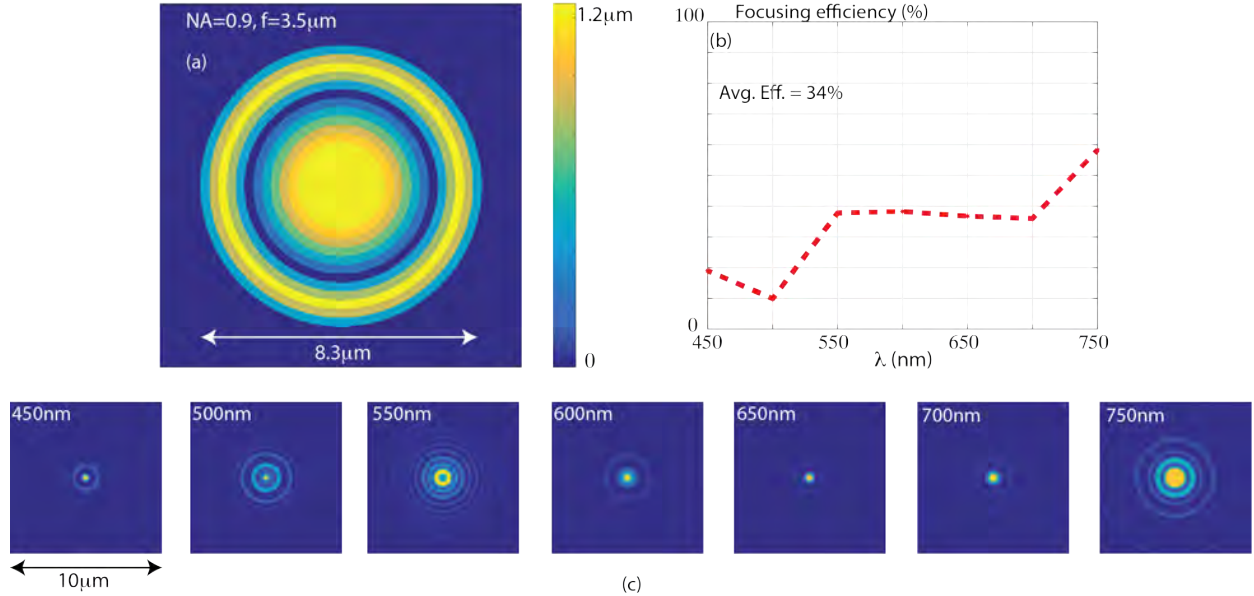

*Figure 4:* *NA=0.9 diffractive lens. (a) Design details. (b) Simulated focusing efficiency as a function of wavelength. (c) Simulated focal spots at various design wavelengths in the visible spectrum.*

We show here that planar diffractive lenses, when designed properly and fabricated carefully are sufficient for broadband imaging. By extending the fabrication process to industry-relevant lithographic scales and large area replication via nanoimprinting,[27] it is possible to envision planar lenses enabling imaging with very thin form factors, low weights and low costs. Therefore, we believe that our approach will lead to considerably simpler, thinner and cheaper imaging systems.

**Methods**

The achromatic lenses were patterned on a photoresist film atop a glass wafer using grayscale laser patterning using a Heidelberg Instruments MicroPG101 tool. The exposure dose was varied as a function of position in order to achieve the multiple height levels dictated by the design.

The devices were characterized on an optical bench by illuminating them with broadband collimated light, whose spectral bandwidth could be controlled by a tunable filter. The focus of the lenses were captured on a monochrome CMOS sensor for characterization of the PSF. Imaging performance of the lenses were tested in prototype camerasas described in the main text with various color objects under various illuminations including ambient sunlight.

**References**

1. Kress, B., and Meyrueis, P. Digital diffractive optics: An introduction to planar diffractive optics and related technology. Digital Diffractive Optics: An Introduction to Planar Diffractive Optics and Related Technology 396. ISBN 0-471-98447-7. Wiley-VCH (2000).

2. Menon, R., Rogge, P. and Tsai, H-Y. Design of diffractive lenses that generate optical nulls without phase singularities. *J. Opt. Soc. Am A* **26**(2), 297-304 (2009).

3. Wan, X., Shen, B. and Menon, R. Diffractive lens design for optimized focusing. *J. Opt. Soc. Am A* **31**(12), B27-B33 (2014).

4. Arbabi, A., Arbabi, E., Kamali, S. M., Horie, Y., Han, S. and Faraon, A. Miniature optical planar camera based on a wide-angle metasurface doublet corrected for monochromatic aberrations. *Nat. Comm.* **7** 13682 (2016).

5. Khorasaninejad, M., Chen, W. T., Devlin, R. C., Oh, J., Zhu, A. Y., and Capasso, F. Metalenses at visible wavelengths: Diffraction-limited focusing and subwavelength resolution imaging. *Science*, **352**(6290), 1190-1194, (2016).

6. Arbabi, A. and Faraon, A. Fundamental limits of ultrathin metasurfaces. *arXiv:1411.2537* [physics.optics] (2016).

7. Aieta, F., Genevet, P., Kats, M. A., Yu, N., Blanchard, R., Gaburro, Z. and Capasso, F. Aberration-Free Ultrathin Flat Lenses and Axicons at Telecom Wavelengths Based on Plasmonic Metasurfaces. *Nano lett.* 12, 4932 (2012).

8. Genevet, P., Yu, N., Aieta, F., Lin, J., Kats, M. A., Blanchard, R., Scully, M. O., Gaburro, Z., and Capasso, F. Ultra-thin plasmonic optical vortex plate based on phase discontinuities. *Appl. Phys. Lett.* **100** 013101 (2012).

9. Aieta, F., Kats, M. A., Genevet, P. and Capasso, F. Multiwavelength achromatic metasurfaces by dispersive phase compensation. *Science* **347** 6228, 1342-1345 (2015).

10. Arbabi, E., Arbabi, A., Kamali, S. M., Horie Y. and Faraon, A. Multiwavelength metasurfaces through spatial multiplexing. *Sci. Rep.* 6:32803 | DOI: 10.1038/srep32803.

11. Karimi, E., Schulz, S. A., De Leon, I., Qassim, H., Upham, J. and Boyd, R. W. Generating optical orbital angular momentum at visible wavelengths using a plasmonic metasurface. *Light: Science & Applications* **3** e167 (2014).

12. Jahani, S. and Jacob, Z. All-dielectric metamaterials. *Nat. Nanotechnol.* **11** 23–36 (2016).

13. Yu, N., and Capasso, F. Flat optics with designer metasurfaces. *Nat. Mater.* **13** 139–50 (2014).

14. Bao, Y., Jiang, Q., Kang, Y., Zhu, X. and Fang, Z. Enhanced performance of multifocal metalens with conic shapes. *Light: Science & Applications* **6**, e17071 (2017).

15. Bao, Y., Zu, S., Liu, W., Zhou, L., Zhu, X. and Fang, Z. Revealing the spin-optics in conic-shaped metasurfaces. *Phys. Rev. B* **95**, 081406(R) (2017).

16. Shen, B., Wang, P., Polson, R. C. and Menon, R. An ultra-high efficiency Metamaterial Polarizer. *Optica* **1**(5) 356-360 (2014).

17. Chao, D., Patel, A., Barwicz, T., Smith, H. I., and Menon, R. Immersion Zone-Plate-Array Lithography. *J. Vac. Sci. Technol. B*, 23(6), 2657-2661 (2005).

18. Kim, G., Domínguez-Caballero, J. A. and Menon, R. Design and analysis of multi-wavelength diffractive optics. *Opt. Exp.* **20**(3) 2814-2823 (2012).

19. Mohammad, Wang, P., Friedman, D. J. and Menon, R. Enhancing photovoltaic output power by 3-band spectrum-splitting and concentration using a diffractive micro-optic. *Opt. Exp.* **22** (106) A1519-A1525 (2014).

20. N. Mohammad, M. Schulz, P. Wang, and R. Menon. "Outdoor measurements of a photovoltaic system using diffractive spectrum-splitting and concentration." AIP Advances **6** (9) (2016).

21. Wang, P. and Menon, R. Optical microlithography on oblique and multiplane surfaces using diffractive phase masks. *J. Micro/Nanolith. MEMS MOEMS,* **14**(2), 023507 (2015).

22. Wang, P., Mohammad, N. and Menon, R. Chromatic-aberration-corrected diffractive lenses for ultra-broadband focusing. *Sci. Rep.* **6** 21545 (2016).

23. Peng, Y., Fu, Q., Heide, F., Heidrich, W. The diffractive achromat: full spectrum computational imaging with diffractive optics. *SIGGRAPH* '16, July 24-28, Anaheim, CA (2016).

24. Goodman, J. W. Introduction to Fourier Optics. Roberts and Company (2005).

25. Wang, P., Dominguez-Caballero, J. A., Friedman, D. J. and Menon, R. A new class of multi-bandgap high-efficiency photovoltaics enabled by broadband diffractive optics. *Prog. Photovolt: Res. Appl.* **23** 1073–1079 (2015).

26. Khorasaninejad, M., Shi, Z., Zhu, A. Y., Chen, W. T., Sanjeev, V., Zaidi, A. and Capasso, F. Achromatic Metalens over 60 nm Bandwidth in the Visible and Metalens with Reverse Chromatic Dispersion. *Nano Lett.* 17(3) 1819-1824 (2017).

27. Guo, L. J. Recent progress in nanoimprint technology and its applications. *J. Phys. D: Appl. Phys.* **37** R123-R141 (2004).

**Acknowledgements:** We thank Brad Sohnlein and Husain Imam from NKT Photonics for assistance with the super-continuum source. We thank Ganghun Kim for help with blind deconvolution of the Air Force target images. We also thank Patrick Clifford (http://www.pcedits.com) for editing the supplementary videos. We gratefully acknowledge support from a DOE Sunshot Grant, EE0005959, a NASA Early Stage Innovations Grant, NNX14AB13G, and the Utah Science Technology and Research (USTAR) Initiative.

**Author contributions**

NM assembled the optical system, performed all experiments and analyzed the data. MM fabricated the lenses, characterized their geometry, performed all experiments and analyzed the data. BS performed the FDTD simulations. RM conceived and designed the experiments. All authors edited the manuscript.

**Additional information**

Correspondence and requests for materials should be addressed to Rajesh Menon.

**Competing financial interests**

None.

# Supplementary Information

## Broadband imaging with one planar diffractive lens

Nabil Mohammad,[1] Monjurul Meem,[1] Bing Shen,[2] and Rajesh Menon[1,*]

[1] Department of Electrical and Computer Engineering, University of Utah, Salt Lake City UT 84112
[2] MACOM Technology Solutions, NY

*Corresponding author: rmenon@eng.utah.edu

1.  **Design methodology**

Our broadband lens is comprised of concentric rings of varying heights as illustrated in Fig. 1(a). The widths of the rings may be varying as well. In this manuscript, the widths of the rings were kept the same. The diameter of the lens is determined by the numerical aperture of the design, focal length and the longest wavelength of operation. The field in the focal plane is computed using scalar diffraction theory at each wavelength and the corresponding focusing efficiency is also computed. The focusing efficiency is defined as the power focused within 3 X FWHM divided by the total incident power at a given wavelength. The goal of our optimization based design is to maximize the focusing efficiency averaged over all the wavelengths of interest by varying the heights of the rings comprising the lens. We refer to this as the figure of merit (FOM). We utilized the modified binary-search algorithm to perform this optimization (see flowchart below). At first, an initial guess of height distribution is generated (usually a random distribution). In one iteration, all ring-heights are perturbed in a pre-designed manner (a random sequence). A positive unit perturbation (+Δ$h$) is tried. If the updated FOM is increased, then this perturbation is kept, otherwise a negative unit perturbation (-Δ$h$) is applied to this groove. If the new FOM is calculated to increase, then this negative perturbation is kept, otherwise it proceeds to the next groove. The guessed height distribution is updated accordingly. One iteration stops when all grooves are traversed. Termination conditions guarantee convergence, such as a maximum number of total iterations or a minimum FOM improvement threshold between two iterations.

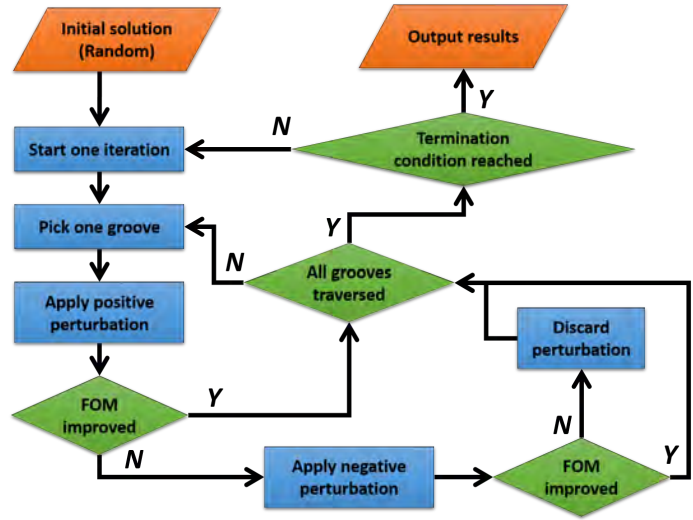

The key design parameters are described below.

*Table S1: Design parameters.*

| NA | f | Ring width | Max. ring height | Number of gray-levels |
|---|---|---|---|---|
| 0.05 | 1mm | 3μm | 2.4μm | 100 |
| 0.18 | 1mm | 1.2μm | 2.6μm | 100 |

## 2. Simulation of focusing performance

The simulated point-spread functions (focal spots) at the design wavelengths for the 2 lenses are summarized in Fig. S1. The measured focal spots (from fig. 1 of the main text) are also included for comparison. It can be seen that the measurements agree well with the simulations. These simulations were performed using scalar diffraction theory.

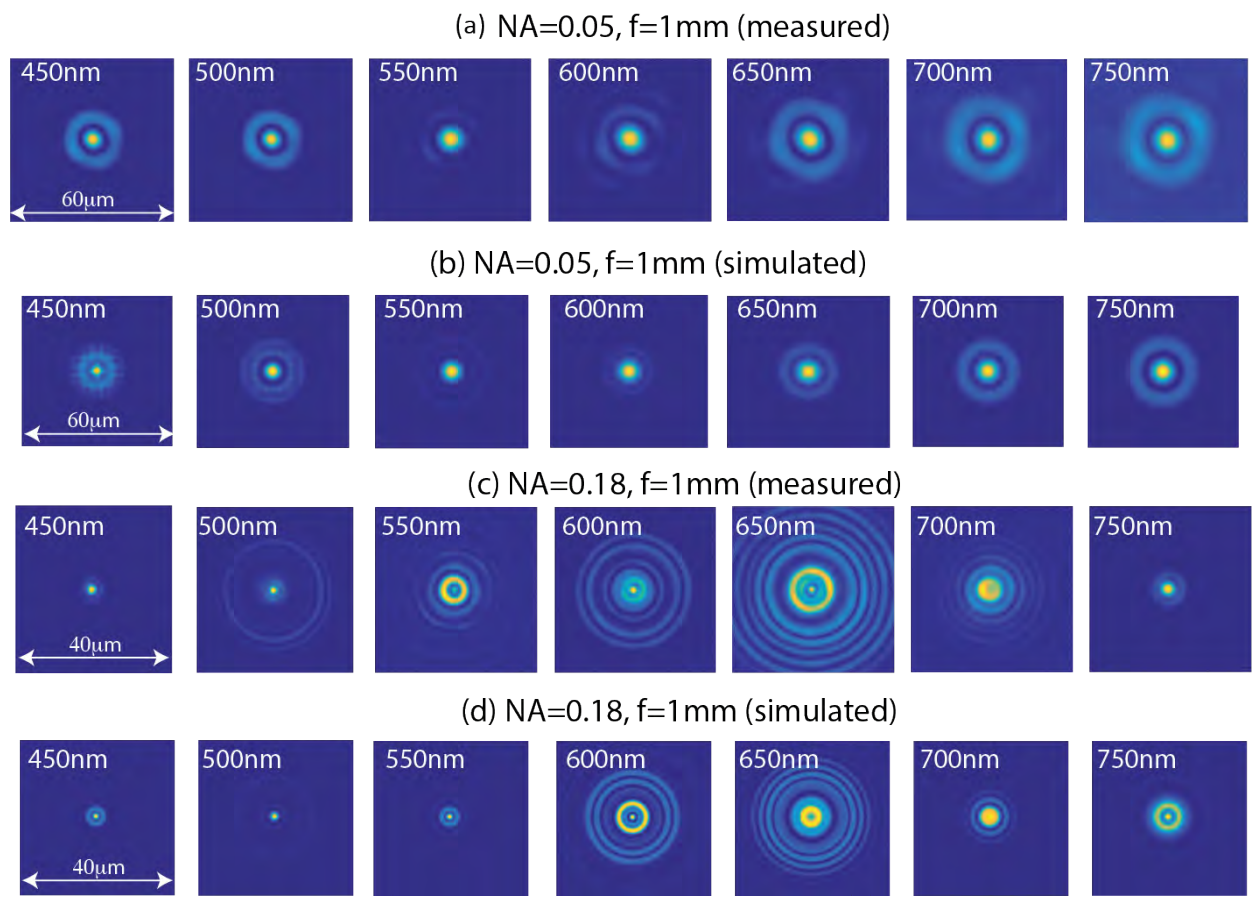

***Figure S1:*** *Focal spots of NA=0.05 lens (a) measured, (b) simulated and those of NA=0.18 lens (c) measured and (d) simulated.*

The simulated focusing efficiency spectra of the 2 designed lenses are shown in Fig. S2. The simulated average efficiency is somewhat higher than the measured ones. The reasons for this is not clear at the moment and we strongly suspect the impact of fabrication errors as described below.

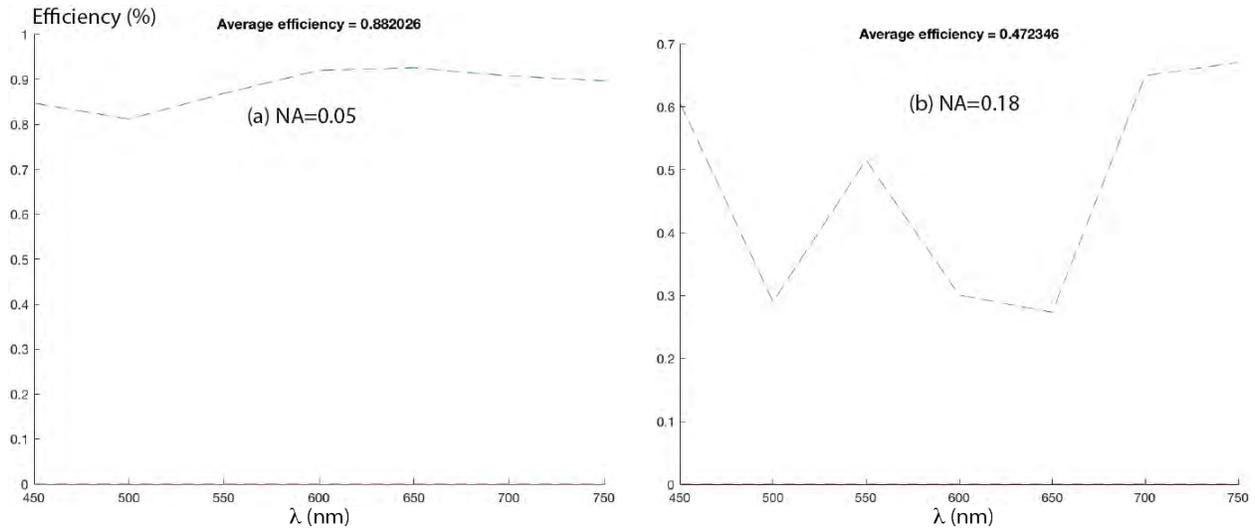

***Figure S2:*** *Simulated focusing efficiency spectrum for (a) NA=0.05 and (b) NA=0.18 lens.*

### 3. Impact of fabrication errors

We measured the heights of the fabricated design (NA=0.18) over 10 randomly selected rings and these are summarized in the top chart in Fig. S3 along with the corresponding ideal design heights. The estimated error has a mean of 968nm and standard deviation of 156nm. Using this information, we simulated the focusing efficiency for 3 random realizations, where the pixel height error was randomly drawn from an uniform distribution with the same mean and standard deviation as the measured values. The results are summarized in the bottom panel of Fig. S3 and provide more proof for our hypothesis that the fabrication errors are likely reason for the reduced focusing efficiency. Another possibility, which requires additional study is the surface roughness of the photoresist after development.

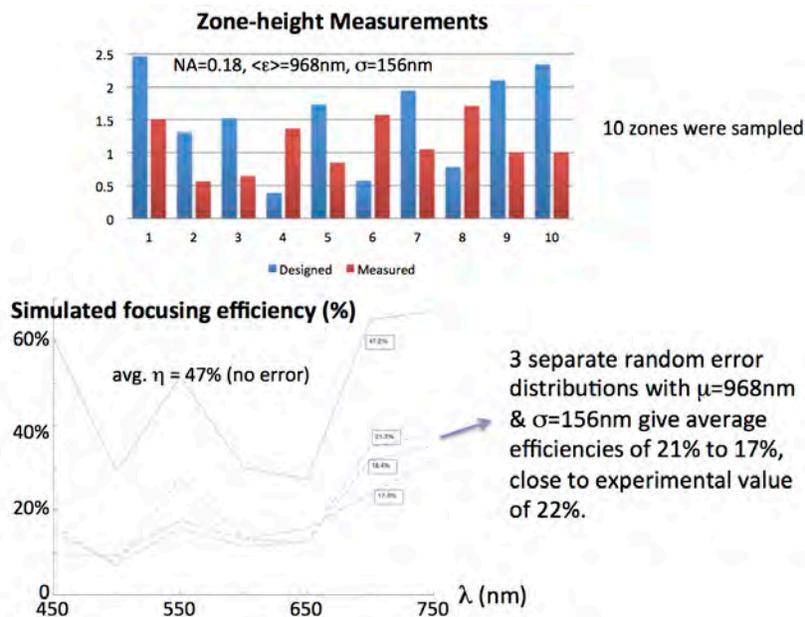

*Figure S3:* Top: measured and designed zone heights for 10 randomly selected zones in the NA=0.18 lens. Bottom: Impact of errors in the zone heights on focusing efficiency.

**4. Focal spot characterization setup**

The flat lenses were illuminated with expanded and collimated beam from a SuperK EXTREME EXW-6 source (NKT Photonics) and the SuperK VARIA filter (NKT Photonics). The wavelength and bandwidth can be changed using the VARIA filter. The focal planes of the flat lenses were magnified using an objective (RMS20X-PF, Thorlabs) and tube lens (ITL200, Thorlabs) and imaged onto a monochrome sensor (DMM 27UP031-ML, Imaging Source). The setup is shown in Fig. S4. Here, f represents the focal length of the flat lens and w.d. (roughly 2mm) is the working distance of the objective. The gap between objective and tube lens was ~90 mm and that between the sensor and the backside of tube lens was about 148mm. The magnification of the objective-tube lens was 22.22X.

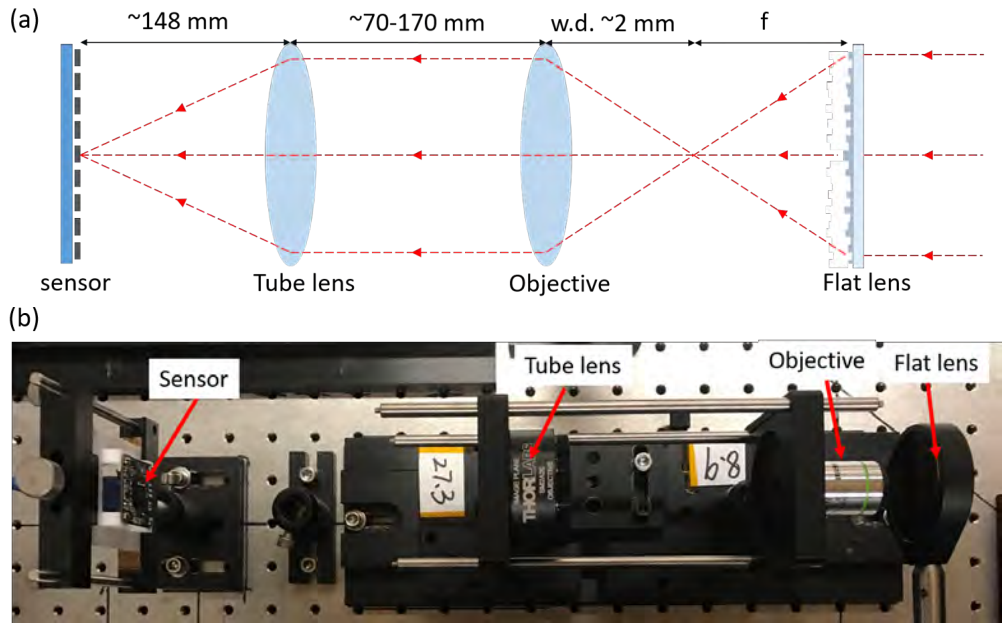

*Fig. S4:* (a) Schematic of the setup for focal spot characterization. (b) Photograph of the setup.

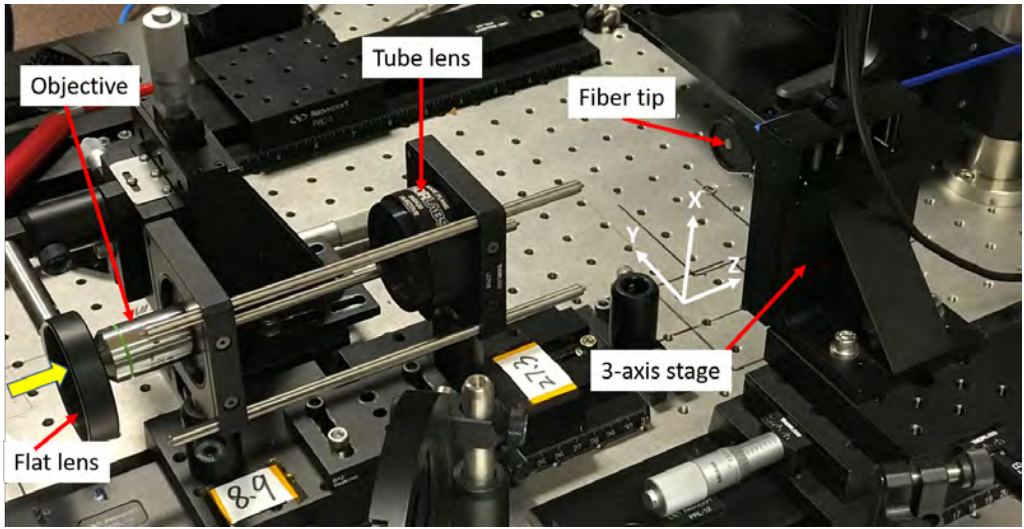

*Fig. S5: Schematic of the setup for focusing efficiency measurement.*

To experimentally determine the focusing efficiency, we used the same setup but now we replaced the monochrome sensor with a 400 µm core diameter fiber tip (P400-1-UV-VIS, Ocean Optics) which in turns was connected to a spectrometer (Jaz Spectrometer, Ocean Optics). The setup is shown in Fig. S5. The flat lens was illuminated with expanded and collimated beam from the SuperK (445nm to 755nm). The fiber tip was scanned in X and Y directions using motorized stages so that the fiber tip was aligned with the peak of the magnified focal spot of the flat lens. A slight adjustment in the Z direction was also made to ensure that the integrated signal on the spectroscope was maximum. A reference signal was recorded with light passing through the unpatterned photoresist. Focusing efficiency was then calculated using the following equation: Focusing efficiency = (spectrometer signal when fiber tip aligned to the peak of magnified psf) / [(reference signal) X (area of magnified lens aperture) / (area of fiber tip aperture)]

### 5. Imaging setup

For imaging experiment, the 1951 USAF resolution test chart (R3L3S1N, Thorlabs) was used as the object. The flat lenses were used for imaging the object on to the sensor. A diffuser was placed behind the USAF target. The experimental setup is shown in Fig. S6. The USAF target was illuminated with the design wavelengths with 10nm bandwidth and corresponding images were captured using a monochrome sensor (DMM 27UP031-ML, Imaging Source). The exposure time was adjusted to ensure that the images did not get saturated. In each case, a dark frame was recorded and subtracted from the USAF target images.

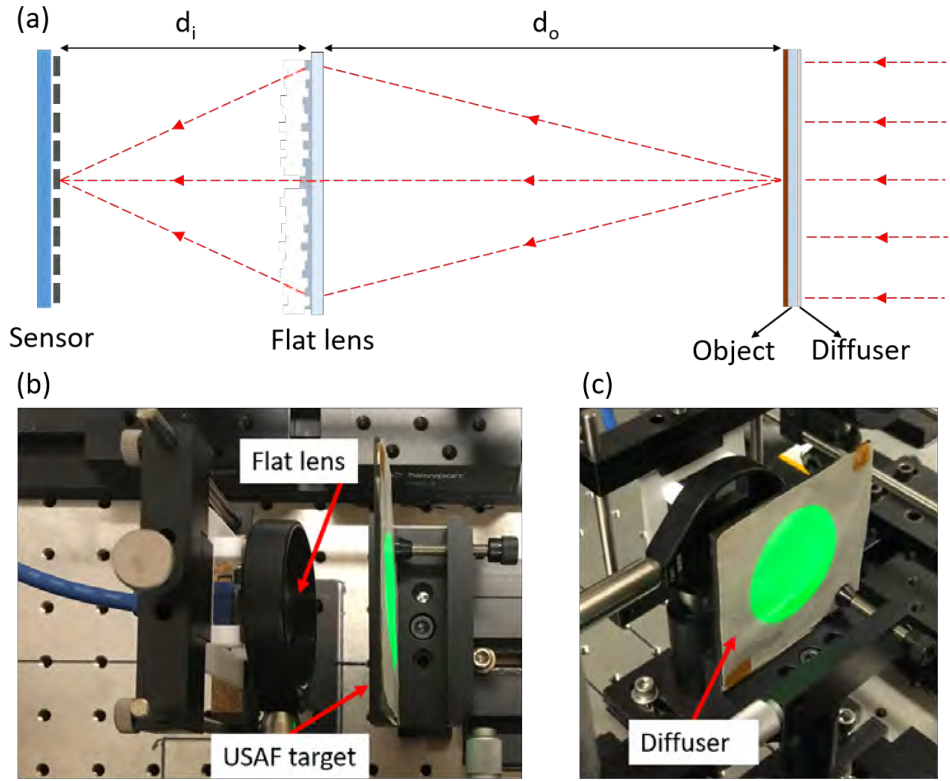

***Fig. S6:*** *(a) Schematic of the imaging setup. (b) and (c) Photographs of the setup.*

***Table S2****: Focal length, magnification, object and image distance*

| Lens NA | Object distance (mm) | Image distance (mm) | Focal length (mm) | Magnification |
|---|---|---|---|---|
| 0.05 | 42.3 | 1.0242 | 1 | 0.0242 |
| 0.18 | 38.8 | 1.0265 | 1 | 0.0264 |

**6. Color camera setup**

Fig. S7 shows the photographs of the color camera and its components. The color camera consists of a color sensor (DFM 72BUC02-ML), the flat lens and an IR-cut filter. These components were put onto a cage mount which was placed inside a 3D printed enclosure. The IR-cut filter was used only for outdoor photography.

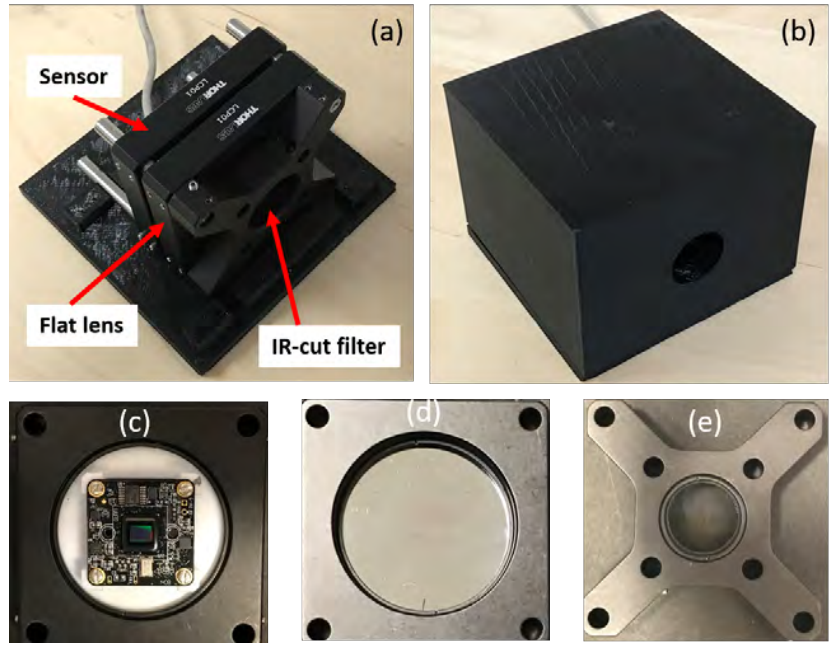

***Fig. S7:*** *(a) Photograph of the color camera. (b) a 3D printed enclosure. Components of the camera: (c) color sensor, (d) flat lens and (e) IR-cut filter.*